\documentclass{article}
\usepackage{graphicx} 
\usepackage{amsmath}
\usepackage{mathtools} 
\usepackage{nicefrac}
\usepackage{xcolor}
\usepackage[labelfont=bf]{caption}

\newcommand{\BF}[1]{\boldsymbol{#1}}
\newcommand{\sub}[2]{#1_{\mathrm{#2}}}
\newcommand{\super}[2]{#1^{\mathrm{#2}}}

\newcommand{\BTBF}{\BF{B}^{\mathrm{T}}}

\newcommand{\CeffBF}{\boldsymbol C^{\mathrm{eff}}}

\newcommand{\sinksum}{\sum_{j \in \mathrm{sinks}}}

\title{Modeling Full-Scale Leaf Venation Networks}
\author{Lars Erik J. Skjegstad \& Julius B. Kirkegaard}
\date{\today}

\begin{document}

\maketitle

\textbf{Abstract:}
The vascular network of leaves, comprising xylem and phloem, is a highly optimized system for the delivery of water, nutrients, and sugars.
The design rules for these naturally occurring networks have been studied since the time of Leonardo da Vinci, who constructed a local rule for comparing the widths of in- and outgoing veins at branch points.
Recently, physical models have been developed that seek to explain the full morphogenesis of leaf venial networks in which veins grow in response to local hydrodynamic feedback.
Although these models go beyond simple local rules, they are challenging to compare to experimental data.
Here, we extend these hydrodynamic models to a state where the direct comparison with images of full leaves becomes possible on the level of individual veins.
We present a dataset of the venial networks of leaves that maintain full network topology and use this to discuss the benefits and drawbacks of such a direct comparison.
We apply our approach to the direct estimation of a sink fluctuation parameter, demonstrating consistency within distinct leaf species.
Finally, we utilize the ability of the model to run on full leaves to define and calculate exponents for a Murray's law that applies to reticulate venation networks.

\section{Introduction}
\begin{figure}
    \centering
    \includegraphics[angle=90, width=1.0\linewidth]{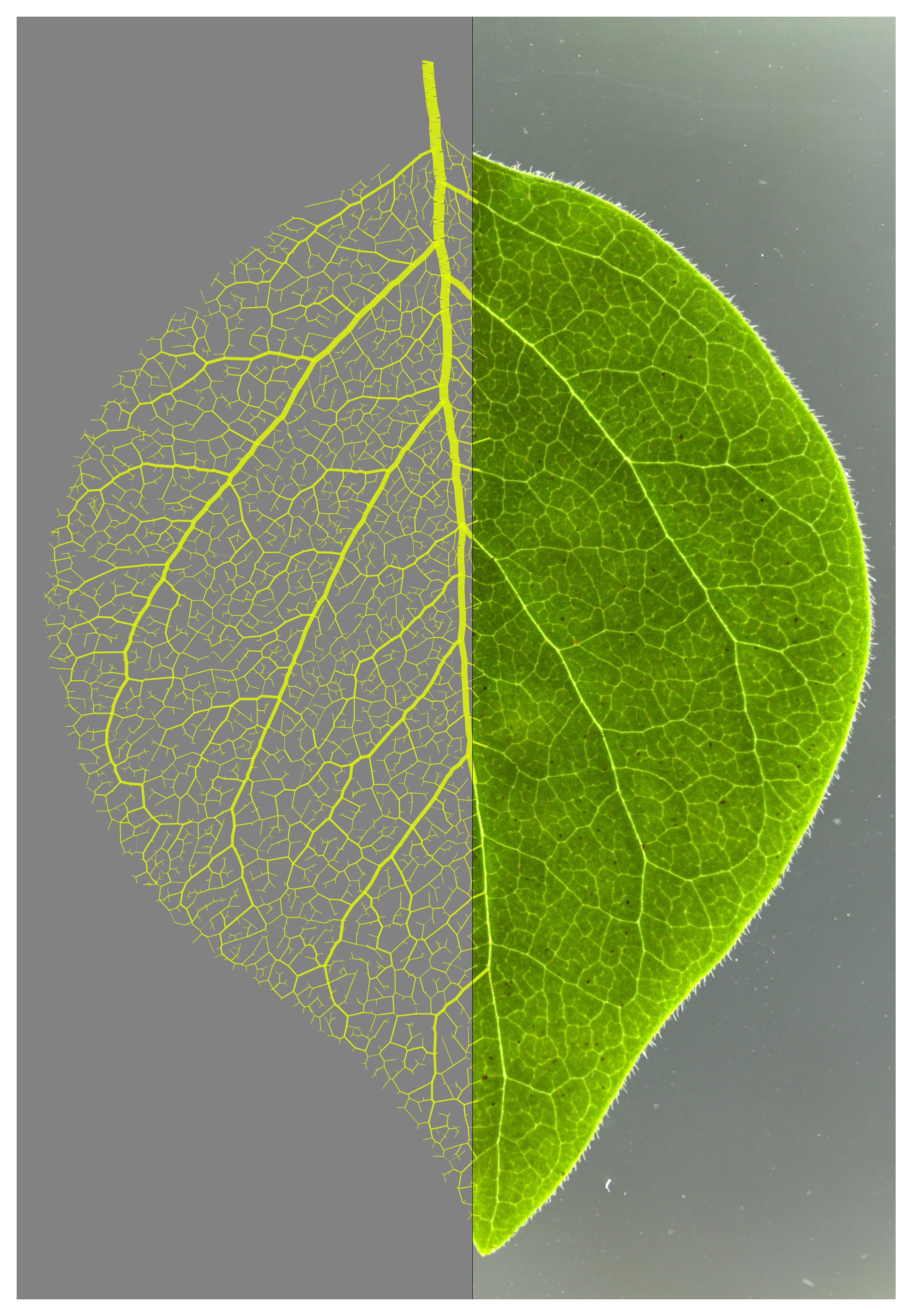}
    \caption{Sample image of \textit{S. albus} (top half) with corresponding model output venation pattern (bottom half).}
    \label{fig:division}
\end{figure}
Leaf venation networks are intricate structures that are key to the overall health, growth, and adaptability of plants \cite{sack2013leaf, katifori2018transport}.
These veins are comprised of the xylem and phloem tissues.
The xylem is primarily responsible for transporting water and minerals absorbed from the soil out to the leaves, where photosynthesis occurs, while the phloem distributes the photosynthetic products, such as sugars and other organic molecules, back throughout the plant.
Additionally, the venial network can function as structural support for the leaf.

The venation structure typically exhibits self-similarity, dominated at the high level by a tree-like topology.
However, the structure becomes more complex in environments where leaves are susceptible to damage, or where local changes to the demand of water within the leaves occur due to, e.g., variations in solar radiation.
This complexity is characterized by the emergence of network loops, which are redundant structures that serve as adaptive mechanisms for the leaf to maintain functionality and resilience under changing or adverse conditions.
These structural features ensure that the network can continue to satisfy the physiological demands of the leaves, even when external factors introduce variability.

Historically, one of the earliest documented explorations in leaf venation was made by Leonardo da Vinci \cite{da2012notebooks}.
His observation that the total cross-sectional area of all branches in a tree remains consistent from the base trunk to the leaves foreshadowed what is now known as Murray's law \cite{murray1926physiological}.
Since then, the understanding of leaf venation networks has advanced significantly on two distinct fronts.
First, automated image analysis and network analysis have laid the foundation for studies on the statistical properties and topology of vein networks.
Second, hydrodynamic models for transport networks have been developed,
explaining not only the theory behind the local variations in vein thickness but also the overall morphogenesis of the network structure, particularly in small idealized networks.
In this study, we bridge these two approaches by integrating automated image analysis and hydrodynamic modeling techniques.
We develop methods to apply these theoretical models to networks extracted from images of actual leaves,
allowing us to examine how the principles derived from the idealized models apply to complex, naturally occurring venation patterns [Fig. \ref{fig:division}].

In a seminal paper, Bohn \& Magnasco \cite{bohn2007structure} proposed a general framework for investigating the properties of optimal transport networks, which has been applied to a diverse set of problems, including leaves \cite{ronellenfitsch2016global, katifori2012quantifying}, river networks \cite{konkol2022interplay}, vascular blood systems \cite{katifori2012quantifying, kirkegaard2020optimal}, and traffic flow \cite{lonardi2021designing, lonardi2023bilevel}.
Their hydrodynamic model shows that in steady-state systems,
i.e., where the current is constant,
the optimal network configuration adopts a tree-like topology characterized by a complete absence of loops,
which is highly different from the networks seen in many types of leaves, where loops are prevalent.
However, when the system is subjected to fluctuating loads or probabilistic edge failures, redundancies become necessary,
which leads to the emergence of more realistic reticulate networks as optimal structures \cite{katifori2010damage, corson2010fluctuations, waszkiewicz2024goldilocks}.

The hydrodynamic model of network optimality \cite{bohn2007structure} suffers from the existence of many local optima, and local adaptation rules \cite{hu2013adaptation} thus rarely lead to the globally optimal network.
By taking into account the growth of the leaf \cite{ronellenfitsch2016global}, however, more optimal solutions can be reached from local rules, resulting in networks that better compare to real leaves.
In its various formulations, the hydrodynamic model thus qualitatively captures many characteristics of leaf venation. However, two significant challenges arise when analyzing networks at realistic scales.
First, the hydrodynamic model relies on graph-based computations that require the inversion of matrices whose dimensions correspond to the number of nodes or edges in the network.
For optimization processes that necessitate a large number of iterations, this matrix inversion becomes computationally prohibitive in large-scale systems.
Second, the number of local optima increases with network size, complicating the analysis of global optimality.
To address these challenges, we employ two strategies to enable computations on full-size networks:
(1) we partition the full-size networks into smaller subsections and optimize these substructures separately while preserving the influence of the surrounding network,
and (2) we mitigate the issue of the numerous local optima by restricting our focus to the hydrodynamic optimality only in the vicinity of the network topology observed in the relevant leaf.

We demonstrate the utility of our model by applying it to images of leaves, showcasing the consistency and variability within leaf species by using several samples from each species.
In particular, we exemplify on \textit{Symphoricarpos albus}, \textit{Lonicera xylosteum}, and \textit{Crataegus monogyna}, which are native to the surroundings of the University of Copenhagen.
We develop a hydrodynamic model with fluctuating sinks and evaluate it on networks extracted from the entire leaf, which results in data on each individual network node and edge.
We subsequently compare the model output vein widths with those from the image data, discussing the possible reasons for discrepancies between them, and use the data to perform parameter estimations.
Lastly, we do an empirical study of Murray's law on the full-scale ``perfect'' leaves that result from our simulations.
We suggest an immediate extension to Murray's law to account for the reticulate structure in the networks and investigate the limits of applying such a law in the presence of perfect data.

\begin{figure}[h]
    \centering
    \includegraphics[width=1.0\linewidth]{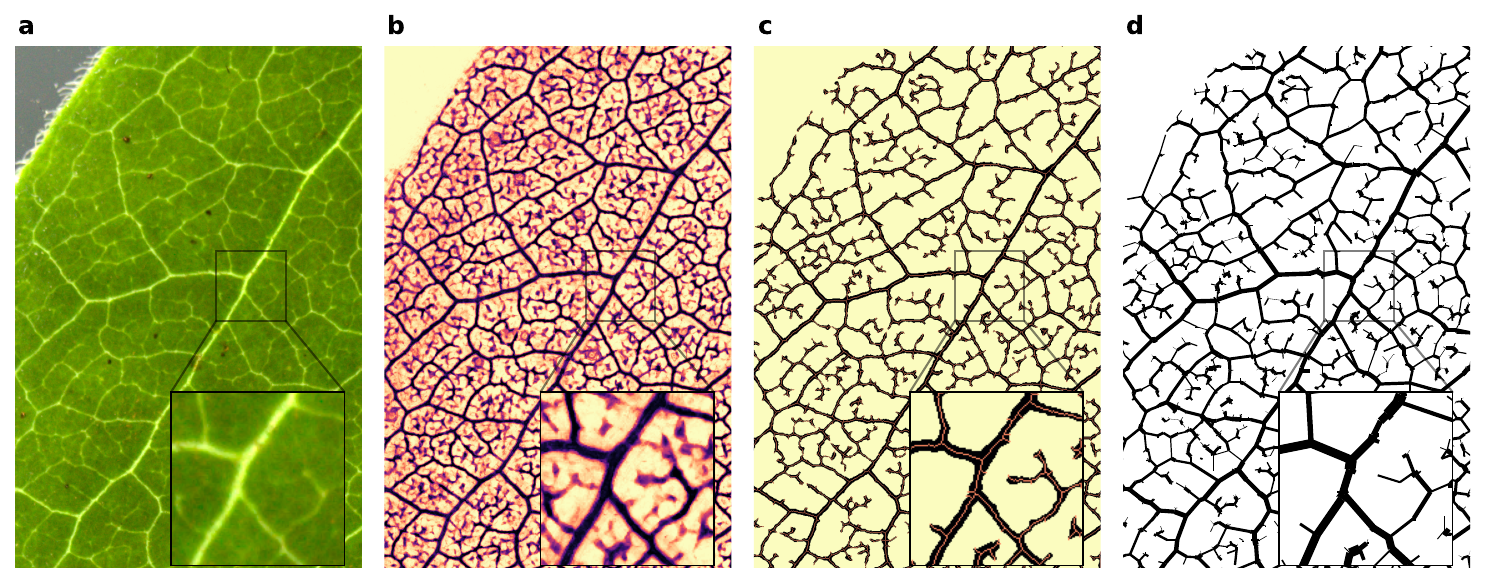}
    \caption{Network extraction. \textbf{(a)} Section of input image. \textbf{(b)} Semantic image segmentation of the veins from background. \textbf{(c)} Binary segmentation and the network skeleton. \textbf{(d)} The final graph extracted, represented by the coordinates and incidences of all nodes and edges, as well as the lengths and widths of each edge.}
    \label{fig:network_creation}
\end{figure}

\section{Model}
We are interested in comparing theoretical models on the formation of complete venial networks in leaves directly to leaf data, a process which relies on the availability of networks describing these full-scale systems.
To that end, we employ a custom pipeline for extracting the network topology and edge attributes from leaf images.
The process, illustrated in Fig. \ref{fig:network_creation}a-d, relies on the deep learning segmentation of veins \cite{ilastik}, as well as a custom edge extraction algorithm.
Details can be found in the Supplemental Information and the associated source code.

The network structure is represented by a directed, planar graph consisting of a total of
$\sub{n}{n}$ nodes and $\sub{n}{e}$ edges, each edge being associated with a length $\ell_i$ and width $w_i$.
The venial network has two roles:
it transports and distributes fluid from the petiole to the rest of the leaf (xylem)
and it distributes the photosynthetic products from the leaf cells to the plant (phloem).
Mathematically, this leads to dual formulations with identical end results, and here we formulate the problem in terms of xylem only.
Thus, to model the fluid flow through the network we place a fluid source at the petiole, and a sink at every node in the network that accounts for both fluid evaporation and photosynthetic consumption.

Assuming laminar fluid flow, we model each edge as a cylindrical pipe, such that the flux $f_i$ over an edge $i$ is given by the Hagen-Poiseuille law,
\begin{equation}
    f_{i} = \frac{c_i}{\ell_i} \, (\Delta p)_i,
\label{eq:poiseuille}
\end{equation}
where $\Delta p$ is the pressure difference between the two nodes connected by the edge, and $c$ is the conductivity of the edge, which obeys proportionality $c \sim w^4$.
We further define the conductance $\super{c}{eff} = c / \ell$.

Defining $\BF{B}$ as the $\sub{n}{n} \times \sub{n}{e}$ incidence matrix for the network allows us to write Eq. \eqref{eq:poiseuille} for the entire network as
\begin{equation}
    \BF{f} = \CeffBF \BTBF \BF{p},
\label{eq:flows_matrix}
\end{equation}
where $\BF{f}$ and $\BF{p}$ are flux and pressure vectors, respectively,
and $\CeffBF$ is a matrix with the conductances on its diagonal.

Combining this with Kirchhoff's law
\begin{equation}
    \BF{B} \BF{f} = \BF{s},
\label{eq:kirchhoff}
\end{equation}
where $\BF{s}$ is the source-sink vector, gives us
\begin{equation}
    \BF{B} \CeffBF \BTBF \BF{p} =\BF{A} \BF{p} = \BF{s},
\end{equation}
where $\BF{A} = \BF{B} \CeffBF \BTBF$ is the weighted graph Laplacian, a symmetric matrix of rank $\sub{n}{n} - 1$.
Since $\BF{s}$ is in the column space of $\BF{A}$, we can solve for the pressures as
\begin{equation}
    \BF{p} = \BF{A}^{\dag} \BF{s},
\label{eq:pressures}
\end{equation}
where $\dagger$ denotes the pseudo-inverse (see SI for an efficient computation of the above).
Lastly, inserting Eq. \ref{eq:pressures} into Eq. \ref{eq:flows_matrix}, we can compute the fluxes as
\begin{equation}
    \BF{f} = \CeffBF \BTBF
    \BF{A}^{\dag} \BF{s} \equiv \BF{Gs},
\label{eq:flows_final}
\end{equation}
where we have defined the matrix $\BF{G}$ for convenience.

\vspace{1em}

We focus our attention on energy efficient networks, which have previously been shown to be a suitable model to describe leaf venial systems \cite{bohn2007structure, katifori2010damage, corson2010fluctuations, hu2013adaptation, ronellenfitsch2016global}.
Thus we minimize the energy dissipation
\begin{equation} \label{eq:power}
    P = \sum_i \frac{f^2_i}{\super{C}{eff}_{ii}},
\end{equation}
subject to constant material cost $\sum_i \ell_i c^{\gamma}_i$.
Here $\gamma$ is a parameter that determines the cost
of building and maintaining the (biological) network.
The higher the value of $\gamma$, the flatter the hierarchy of vein widths,
as thicker widths become more costly.
Using the method of Lagrange multipliers, it can be shown that the optimal conductivities must scale with the fluxes as \cite{bohn2007structure}
\begin{equation}
    c_i \propto | f_i |^{2 / (1 + \gamma)}.
\label{eq:scaling}
\end{equation}
This equation can henceforth be used to find the optimal conductivities by fixed-point iteration.

\vspace{1em}
We have now fully specified the theoretical foundation for the calculation of fluxes and optimal conductivities for a given source-sink vector $\BF{s}$.
A standard choice is $\BF{s} = [1, -\frac{1}{n - 1}, -\frac{1}{n - 1}, \cdots, -\frac{1}{n - 1}]$, where each sink is constant and of equal magnitude.
To enable computations on the non-uniform node distribution that emerges from leaf-image network extraction,
we assign to each node $j$ a source or a sink term
proportional to the area $a_j$ of the Voronoi cell associated with the node.
This is motivated by the assumption that the drainage around a node is proportional to the surface area surrounding that node.

Power minimization with a constant source-sink vector $\BF{s}$ leads to networks with tree topologies \cite{bohn2007structure}.
However, the loopless optimal structures emerging from this are vastly different from the highly loopy networks seen in many species of leaves.
Several models have been developed to mitigate this issue, e.g., the cut-bond and the moving sink models \cite{katifori2010damage, corson2010fluctuations}.
Here, we generalize the moving sink model in order to account for the variations in the node-associated drainage area.
In the original model \cite{katifori2010damage}, the squared flux of Eqs. \eqref{eq:power} and \eqref{eq:scaling} is averaged over a distribution of source-sink vectors, parameterized in terms of sink nodes, i.e.,
\begin{equation} \label{eq:averagesink}
    \langle f_i^2 \rangle = \frac{1}{\sub{n}{n} - 1} \sum_k f_i^2(\BF{s}(k)),
\end{equation}
where the sum in $k$ runs over all sink nodes.
Note that the average is over the squares of flows, which is what distinguishes a model using fluctuating sinks from a static sink model.

In our approach, the distribution of sink values is constructed as a combination of constant overall network load and locally fluctuating sink sizes.
Thus we define source-sink vectors with excess fluid removal at the $k$-th node, making sure that the sink sizes are proportional to the corresponding drainage areas:
\begin{equation}
s_j(k) =
    \begin{cases*}
      1                                              & $j = $ source index \\
      -a_j (\sub{s}{avg} + \delta_{jk} \sigma) C(k)       & otherwise.
    \end{cases*}
\end{equation}
Here, $\sub{s}{avg} = \frac{1 - \sigma}{\sub{n}{n} - 1}$ is a static value for all sinks, and $\delta_{jk} \sigma$ represents the influence of the moving sink, the relative amplitude of which is controlled by the sink fluctuation parameter $\sigma \in [0,1]$.
In the limit $\sigma = 0$, which corresponds to no sink fluctuation,
we have that the fluid removal is solely dependent on the drainage area.
Conversely, in the limit $\sigma = 1$, we have full sink fluctuation,
i.e., $s_k = -1$, and $s_{j \neq k} = 0$ (excluding the source node).
This limit corresponds to the original moving sink model \cite{katifori2010damage}.
Lastly, the normalization constant $C(k) = (\sigma a_k + \sub{s}{avg} \sinksum a_j)^{-1}$ ensures that for any $k$, $\sum s_j(k) = 0$, 
and is necessitated by the fact that the amount of fluid entering the leaf
through the source always equals the amount escaping through the sinks.
Keeping the source term constantly equal to unity, and only letting the sink values vary, corresponds to constant load on the system level.

\subsection{Optimization on extracted graphs}
The theoretical model specified above can in principle be run on any connected graph.
Our model uses networks extracted from images of real leaves as input.
This raises some interesting questions.
For example, how can we find the parameters that best reproduce the vein widths observed in leaf data,
and is it possible to define new relevant statistics based on the knowledge of the fluid flow in the venial network?

To enable computations on the large networks of real leaves, we are faced with two complications:
(1) The theoretical model is highly non-convex, leading to issues in choosing which optima to compare to, and (2) It is computationally expensive to run models on networks with the number of nodes and edges in the thousands.

\paragraph{Hardware acceleration}
We mitigate some of the computational bottlenecks by implementing our models in \textsc{jax}, which enables hardware acceleration on GPUs. This allows us to execute large matrix operations far more efficiently than on CPUs, given that GPUs are optimized for such parallel tasks.
The approach does however introduce a trade-off:
While the use of GPUs yields significantly faster computations, it necessitates more careful memory management, as GPUs generally have less memory than CPUs.

\begin{figure}[ht]
    \centering
    \includegraphics[width=1.0\linewidth]{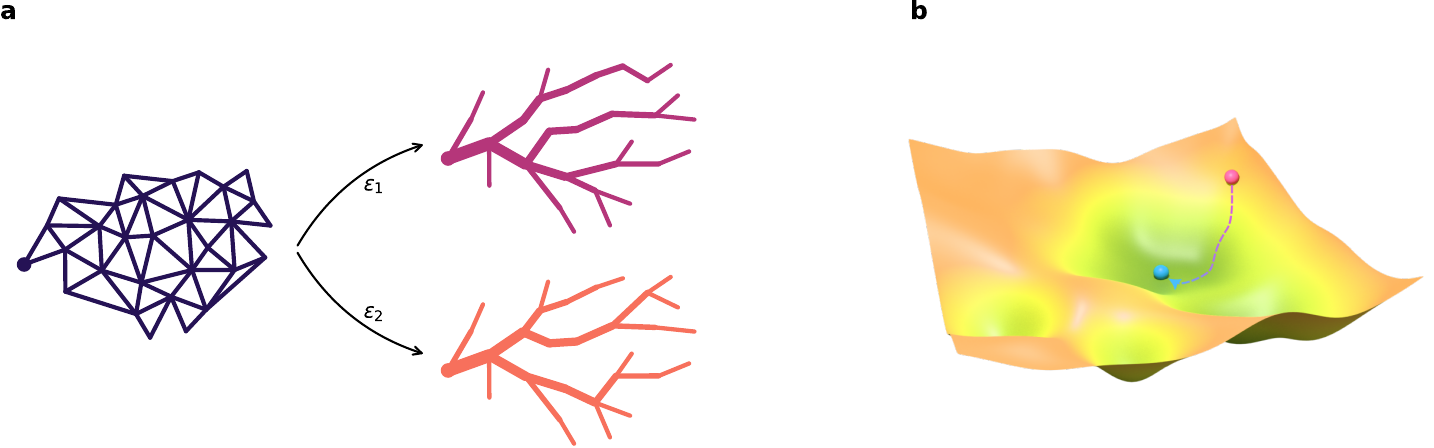}
    \caption{Local optimization.
    \textbf{(a)} Applying two different, but small perturbations, starting from a uniform edge width distribution, leads to very different topologies in the final minima.
    \textbf{(b)} The final solution can be biased by starting near the desired local minimum.}
    \label{fig:local_optimization}
\end{figure}

\paragraph{Non-convexity}
The loss function of a complex transport network model of natural leaves is, in general, highly non-convex \cite{ronellenfitsch2016global}.
As an example of this non-convexity, consider the small system in Fig. \ref{fig:local_optimization}a.
Here, we initialize two networks, each with a uniform distribution of edge widths, with small (but different) perturbations added.
Optimizing the power [Eq. \eqref{eq:power}] results in two entirely different solutions, the networks of which are topologically distinct.
The two minima have very similar power usage, illustrating the highly non-convex nature of the problem.
This makes comparisons with data challenging, as there is a plethora of candidates for the correct local minimum to use.

Our solution to this problem is a simple one:
As our aim is to find the local minimum closest to that of the input data taken from the leaf image, we choose the leaf image data widths as initial values for the iteration process.
This should ideally constitute the best starting point for reaching the desired optimal model output.
The procedure is illustrated in Fig. \ref{fig:local_optimization}b,
where the initial state (pink) is located relatively close to the desired local minimum (blue) in the non-convex landscape.

\paragraph{Subsystems}
Hydrodynamic models of leaf networks are typically run on artificial networks, i.e., small, regular grids \cite{bohn2007structure,katifori2010damage,corson2010fluctuations,hu2013adaptation}.
The size of the networks emerging from the real leaves used for this study is very large, and for most leaves we have $\sub{n}{n}$, $\sub{n}{e} \sim 10^4$--$10^5$.
This sets the scale of the relevant matrices on which we must compute, e.g., the matrix inverse, which scales as $\mathcal{O}(n^3)$.
Clearly, it would be infeasible to perform these computations at every step of an optimization procedure.
We thus adapt our model to these large systems by dividing the systems into subnetworks
and run the model on each section separately.

\begin{figure}[h]
    \centering
    \includegraphics[width=1.0\linewidth]{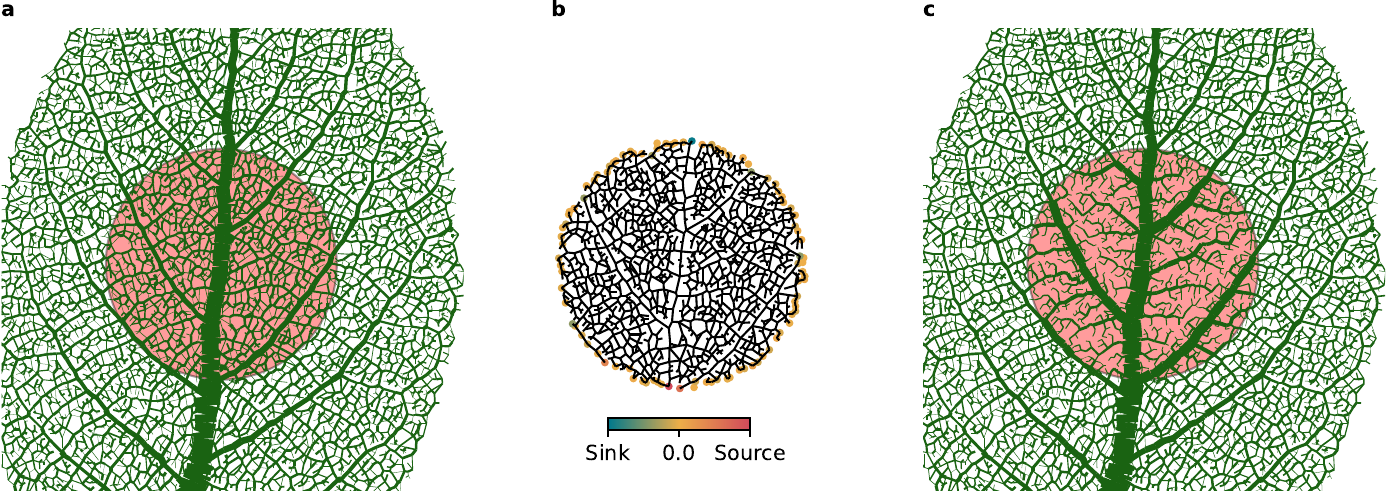}
    \caption{Running the model on a subsystem. \textbf{(a)} Original network with the subnetwork indicated. \textbf{(b)} The subnetwork with redefined sources/sinks, the amplitudes of which are indicated by colors. \textbf{(c)} Output from model with sink fluctuation = 0. Note that most of the reticulate structure has disappeared.}
    \label{fig:substructure}
\end{figure}

Because we initialize the system using the image data widths, we can assume that the changes in a subsystem only affects the surrounding network to a minor degree.
In other words, the effect of a local change has limited range within the entire leaf.
Concretely, we assume that the conductivities outside of a subnetwork remain constant during the optimization of the subnetwork itself.
We use \textit{k}-means clustering to divide the network into a number of subsystems ($\sim 10$), and calculate fluxes by
\begin{equation}
    \BF{\Tilde{f}} = \BF{\Tilde{G}} \BF{\Tilde{s}},
\end{equation}
where the tilde indicates that only the matrix rows and columns corresponding to
the nodes and edges from the subsystem are included.
A source-sink vector of the full system is translated to a source-sink vector for the subsystem by adding the (possibly negative) net flow crossing into the subsystem from the surrounding full system,
\begin{equation}
    \Tilde{s}_j = s_j + \sum_{i \in \mathrm{crossing},j} \super{f}{full}_i,
\end{equation}
where the sum is over the incident edges of node $j$ (if any)
that are also incident to nodes outside of the subsystem.
(Technically, $\Tilde{s}_j$ is calculated using predefined translation matrices that only need to be calculated once for the full system, and do not require updating during optimization -- see SI for a detailed description of these routines.)

An example of this process can be seen in Fig. \ref{fig:substructure}.
The demarcation of one hypothetical subsystem within a full leaf is shown in Fig. \ref{fig:substructure}A,
 and Fig. \ref{fig:substructure}B illustrates how the sink values at the border nodes
are affected by the flow in the full leaf.
Nodes closer to/further from the leaf petiole act as pseudo-sources/-sinks for the subsystem, respectively.
In Fig. \ref{fig:substructure}C the resulting network after performing the optimization on the indicated subsystem is shown,
assuming no sink fluctuation for clarity.
With no sink fluctuation, the stark contrast between the leaf input and model output becomes clearly visible.

\paragraph{The optimal model network}
We now have the necessary ingredients to find the optimal network that most closely models an experimental leaf.
In particular, this enables the direct comparison between the edges from leaf images and model output.
Specifically, this is achieved by solving for the edge conductivities using fixed-point iteration
and then converting them back to edge widths.
We thus initialize $c^{(0)}_i = (\super{w}{data}_i)^4$, and iterate 
\[
        \begin{aligned}
            &f^{(n+1)}_i(k) = \sum_j G_{ij}^{(n)} s_j(k), \\
            &c_i^{(n+1)} = \Bigl \langle \big( f^{(n+1)}_i(k) \big)^2 \Bigr \rangle_k^{1 / (1 + \gamma)}
        \end{aligned}
\]
towards convergence in $\BF{c}$.
The second equation follows from Eq. \eqref{eq:scaling} combined with the fluctuating sink average of Eq. \eqref{eq:averagesink}.
Finally, we convert back to edge widths, $\super{w}{model}_i = c_i^{\nicefrac{1}{4}}$.

We note that in the above, we have worked in arbitrary units of conductivity.
The scale of the conductivity (and of the output widths) is set by the material constraint, which in turn serves to set the scale of the output.
Instead of finding the optimal value for this budget, we simply normalize the output after optimization:
\begin{equation}
            \super{\hat{w}}{model}_i = \frac{\langle \super{w_i}{data} \rangle}{\langle \super{w}{model}_i \rangle} \cdot \super{w_i}{model}.
\end{equation}

\begin{figure}[!h]
    \centering
    \includegraphics[width=1.0\linewidth]{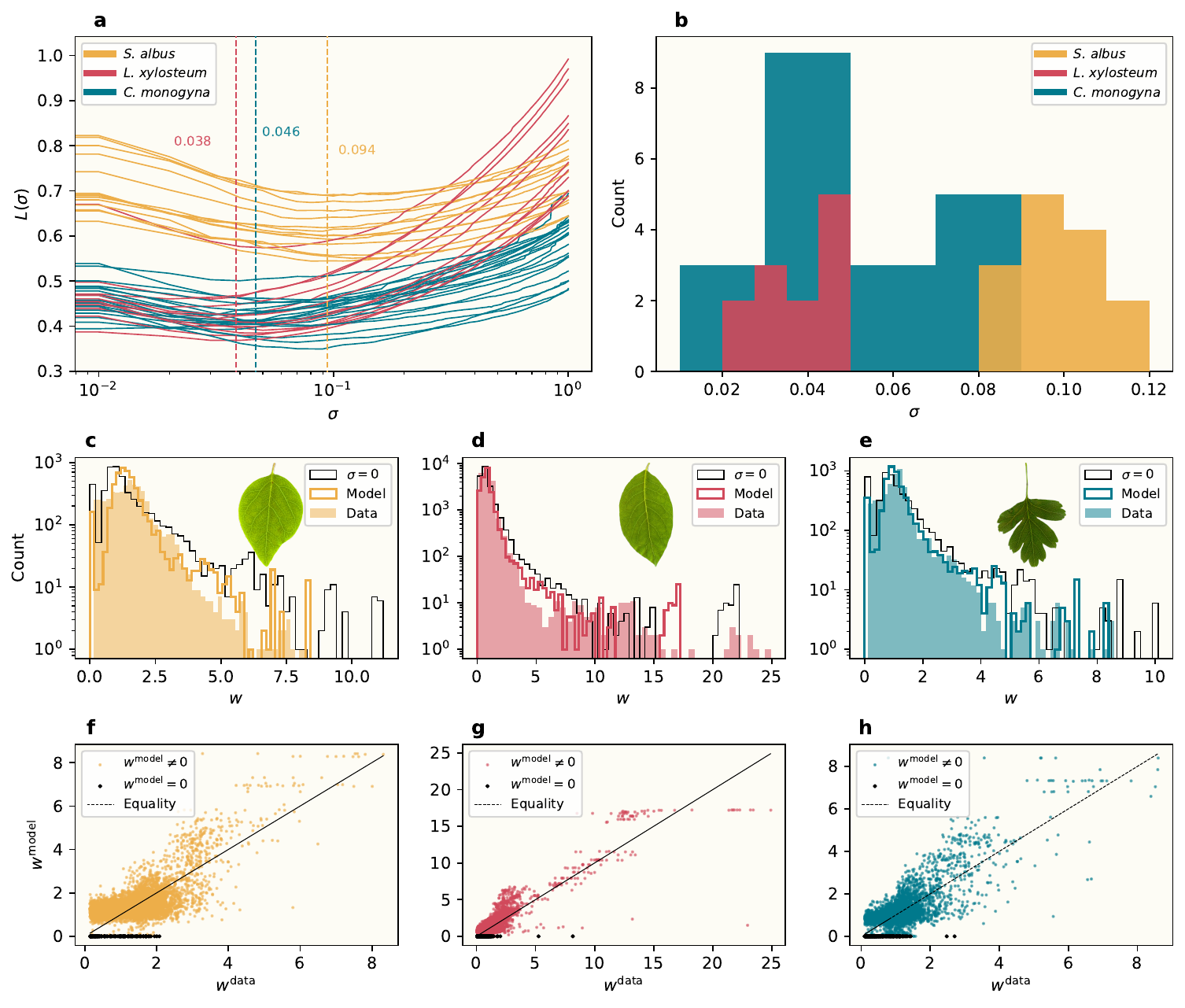}
    \caption{\textbf{(a)} Loss values as a function of sink fluctuation amplitude $\sigma$ for leaves from the three different species.
    The average minimum loss values are indicated for each species.
    \textbf{(b)} Distributions of the optimal values of $\sigma$ for each species.
    \textbf{(c-e)} Vein width distributions for specific leaf samples, one from each species. The distributions are plotted for image data and model output, using the species-specific optimal value of $\sigma$, as well as zero sink fluctuation. Note that the histograms are shown with logarithmic axes. (Insets) The actual leaves analyzed.
    \textbf{(f-h)} Model output vein widths as a function of the corresponding image data vein widths.}
    \label{fig:sigma_fit}
\end{figure}

\section{Applications}
Having fully defined the model, we apply it to a set of leaf images from the three species \textit{S. albus}, \textit{L. xylosteum}, and \textit{C. monogyna}.
For each leaf, we extract a network representation of the observable venation pattern and use this pattern directly as input to our hydrodynamic model to obtain output data on each individual edge.
An example model output is shown in Fig. \ref{fig:division}.
The fact that we can access output data for individual edges allows us to perform a number of experiments:
First, we fit the sink fluctuation amplitude $\sigma$ to the image data from each of the three species.
This will reveal that each species has a distinct optimal value for this parameter.
Second, we investigate the use of Murray's law for reticulate venation networks in ``perfect'' leaves, i.e., leaves that share the graph structure with the corresponding image data leaf, but whose edge widths have been adjusted to obey the theoretical laws.

\subsection{Sink fluctuation amplitude}
We now consider whether direct comparisons of the model output to the extracted leaf networks can be used to fit model parameters.
This contrasts with previous approaches \cite{katifori2012quantifying, ronellenfitsch2015topological}, where summary statistics typically serve as the metrics for comparison.
While this can often be sufficient, the approach ignores the specificities of the individual edges during comparison.
Here, we use the full-scale leaf image network structure as input data, allowing for individual edge-to-edge comparisons with our model.
We hence ask the question of whether this approach can be used to fit theoretical venation models.
Thus, we define a loss function as an average over each individual edge,
\begin{equation}
    L(\sigma) = \Biggl \langle \frac{\left(\super{w}{data}_i - \super{\hat{w}}{model}_i(\sigma) \right)^2}{\super{w}{data}_i} \Biggr \rangle_i,
    \label{eq:loss_func}
\end{equation}
which averages the squared relative difference, weighted by the widths (which corresponds to assigning every vein pixel in the original images equal weight).
We minimize this loss function with respect to the sink fluctuation parameter $\sigma$, while fixing $\gamma = \frac{1}{2}$ (corresponding to a fixed-volume budget).

The results for the three species can be seen in Fig. \ref{fig:sigma_fit}a.
The different leaves within a species exhibit the same qualitative behavior over the range of sink fluctuation values,
and each leaf has a well-defined optimum value for $\sigma$.
Furthermore, the optimal values for a given species are located within a distinct range,
so we use their means (indicated by dashed vertical lines) to represent an optimal value for the species as a whole.

The distributions of the optimal values can be seen in Fig. \ref{fig:sigma_fit}b.
Here we find that both the means and the spreads are markedly different between the species,
indicating variability not only between species but also within species.
The distributions appear to be both distinct and well-defined,
which suggests sink fluctuation as a relevant parameter for species classification based on the venation pattern \cite{ronellenfitsch2015topological}.

\subsection{Individual vein comparisons}
In order to explore how well the model fits the data on an edge-to-edge level, rather than only relying on summary statistics, we now choose one representative leaf from each species,
and study the distributions of the edge widths.
In Figs. \ref{fig:sigma_fit}c--e,
we compare the distributions between the image data and the model output,
using the species-specific fit values as parameters.

The overall shape of the vein width distribution is captured well by the model.
Nonetheless, it is clear that there are discrepancies between the real leaves and model outputs.
These discrepancies are largest for the smallest widths.
In general, the model underestimates the number of small data widths
but conversely produces a significant number of zero-length vein widths.
Note that by construction, there are no zero-length edge widths in the data.
Lastly, a histogram for model output with sink fluctuation set to zero can be seen for all three leaves.
This distribution appears both to have a longer tail and to have a higher number of zero-length widths than the corresponding optimal-sink fluctuation distribution.
This is in line with the fact that the mean vein width is independent of sink fluctuation (due to normalization),
so a higher number of zero-length widths must be compensated by a higher number of wider veins.
The relatively higher number of zero-length widths is also consistent with less sink fluctuation, as expected.

In our approach, we can move beyond the comparison of histograms and compare each individual vein width of the model output with the corresponding vein width in the leaf image data.
This is shown in Figs. \ref{fig:sigma_fit}f--h.
This use of individual edges in the calculations gives a more detailed insight into the model and provides more information than histograms, which describe only the global scale.
This insight reveals that the apparent overlap between data and model does not occur on an individual edge level, but is a global phenomenon.
We can see the same trend in all three leaves:
Although the points are distributed around the line of equality, the varying degrees of spread seem to be significant.
Here we can see clearly how the model overestimates the widths of the smallest veins, for which the model has a preferred size.
This overestimation is partly due to the normalization and compensation for the significant number of model output vein widths with zero size (black dots).
We note that at the level of individual edges, inaccuracies that originate in the preprocessing of the data become more pronounced and that the discrepancies arise due to a combination of model-data mismatch and preprocessing errors.

In conclusion, we find that a single-variable explanation is insufficient and that more complex physical models are needed to fully capture the individual variations in the venation patterns.

\subsection{Murray's law}
\begin{figure}[p]
    \centering
    \includegraphics[width=1.0\linewidth]{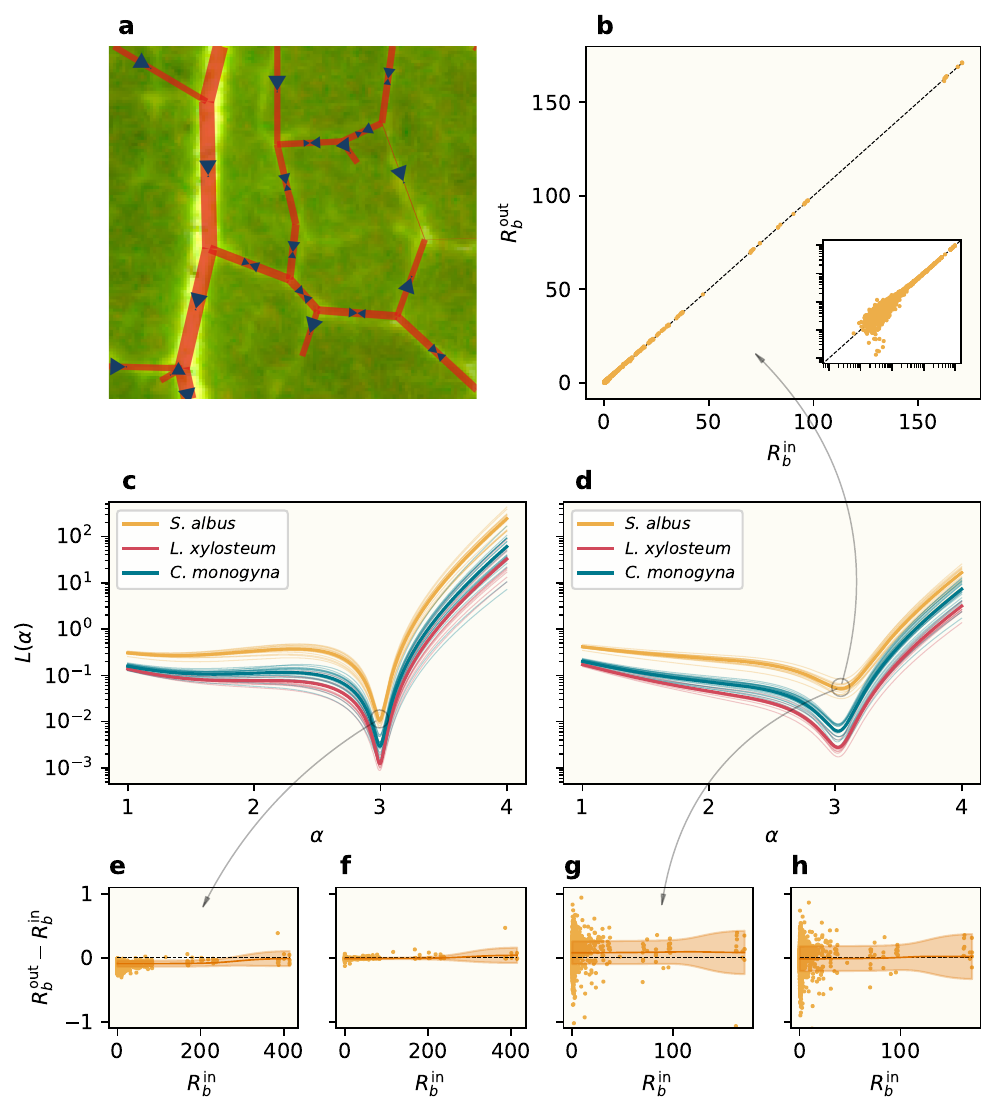}
    \caption{Murray's law for loopy networks.
    \textbf{(a)} Subsection of a leaf image, with model output edge widths (red lines) and fluxes (blue arrows) indicated. The arrow sizes indicate the relative average time the flow moves in the respective directions given by the arrows.
    \textbf{(b)} $\super{R}{out}_b$ vs. $\super{R}{in}_b$ for all nodes in a sample of \textit{S. albus}, for the optimal $\alpha \approx 3.03$ and $\sigma \approx 0.094$.
    The points closely follow the line of equality. (Inset) The same data with both axes logarithmic.
    \textbf{(c-d)} Loss values for individual leaf samples (thin lines), with the mean value for each species (thick lines).
    \textbf{(c)} $\sigma = 0$. The optimal value of $\alpha = 3$ is predicted for all samples.
    \textbf{(d)} $\sigma = 0.094.$ The optimal values have been shifted in the positive direction.
    \textbf{(e-h)}: Residual plots with Gaussian weighted means and standard deviations (SD).
    \textbf{(e-f)} $\sigma = 0$. The residuals are (e) biased negatively, and (f) subsequently corrected by a term proportional to the associated sink area.
    \textbf{(g-h)} $\sigma = 0.094$. The residuals are (g) biased positively, have a larger SD compared to the case of no sink fluctuation, and are (h) corrected by a term proportional to the associated sink area.}
    \label{fig:murray}
\end{figure}

\paragraph{Reticulate network formulation}
We now exploit the fact that we can run the venation model on full-scale leaves to extend Murray's law \cite{murray1926physiological, mcculloh2003water} so that it takes into account the loopy structure in the networks.
Murray's law states that the relationship between the radius $\sub{r}{p}$ of the parent vein
and the radii $\sub{r}{d,\textit{i}}$ of the daughter veins at a branch point is $\sub{r}{p}^\alpha = \sum_i \sub{r}{d,\textit{i}}^\alpha$, with the original law setting $\alpha = 3$, though different values of $\alpha$ have been found in other types of networks \cite{sherman1981connecting, williams2008minimum}.
Notably, the law was originally restricted to tree-topology networks, and attempts at applying it to reticulate networks typically involve simplifying the network structure (such as reducing it to tree topologies) before performing the calculations \cite{katifori2012quantifying, price2013influence}.
To make the model well-defined for loopy networks, we here present a natural extension to it by including sink fluctuation in its definition.

In the following we calculate the Murray sum $\super{R}{in/out}$ for in- or outgoing veins, respectively
(initially ignoring the assumption that the network be leak-free for simplicity).
For a given branch node $b$ with degree $> 2$, we take the sum of in- or outgoing vein radii, each term to the power of the Murray exponent $\alpha$, and then average over possible moving sink positions $k$ in the leaf.
Thus, we define
\begin{equation}
    R_b^{\mathrm{in/out}}(\alpha)
    \equiv
    \Biggl \langle \sum_i \super{I_{b,i}}{in/out}(k) \cdot r_{b,i}^{\alpha} \Biggr \rangle_k,
\end{equation}
where $\super{I_{b,i}}{in/out}(k)$ is an indicator function, the value of which depends on whether the flow of vein $i$ at branch point $b$ is in- or outgoing, given the moving sink position $k$, respectively.
The average over different flow directions is motivated by the fact that the reticulate leaf network has been optimized to accommodate varying flow conditions.
A consequence of this is that the number of terms in the sum can change as the sink distribution changes.

The variations in flux in a venation network due to sink fluctuations are illustrated in Fig. \ref{fig:murray}a.
Here we show a close-up of a leaf image, where the directions of flow are indicated with differently sized arrows, along with the model output edge widths.
The arrow sizes indicate the average fraction of time when the flow moves in a certain direction.
In other words, the flow direction in a given edge depends on the moving sink position $k$.
Consequently, when calculating Murray sums,
we have to account for these changing directions by averaging over all possible sink positions.
The extended version of Murray's law can thus be defined as
\begin{equation} \label{eq:murray}
    R_b^{\mathrm{in}}(\alpha) = R_b^{\mathrm{out}}(\alpha).
\end{equation}
This further allows us to construct a loss function for the system.
Here we take the squared difference of the sums and in turn average the resulting residuals over the number of branch nodes in the leaf,
\begin{equation}
    L(\alpha) = 
    \Biggl \langle \Big(R_b^{\mathrm{out}}(\alpha)
    - R_b^{\mathrm{in}}(\alpha)
    \Big)^2 \Biggr \rangle_b,
    \label{eq:murray_loss}
\end{equation}
the minimum of which defines an optimal value for $\alpha$.

\paragraph{Optimizing $\alpha$}
In the following, we apply the extended Murray's law of Eq. \eqref{eq:murray} to our ``perfect leaves''.
We first search for an optimal value for $\alpha$ for each of the three species of leaf, using the loss function in Eq. \eqref{eq:murray_loss}.
Figs. \ref{fig:murray}c-d show the loss values for a range of different exponents $\alpha$ for sink fluctuation $\sigma = 0$ and its best-fit value (see Fig. \ref{fig:sigma_fit}a), respectively.
The losses are shown for all leaf samples, with the mean loss for each species indicated in bold.
In both cases, we see that the branching observed in the model output leads to well-defined Murray exponents that are consistent on the species level.
With no sink fluctuation, we find optima at $\alpha = 3$, as expected, and with the inclusion of sink fluctuation, all species have optima near $\alpha = 3$ with small variations that are larger between species than within species.
This indicates that the optimal values are significant.

We note that setting $\sigma = 0$ when running the model leads to output networks with tree topologies, for which the original definition of Murray's law should hold exactly.
Indeed, in Fig. \ref{fig:murray}c, we see that the original exponent $\alpha = 3$ is predicted for every individual leaf sample.
However, we also observe that the loss values are not exactly zero for any of the samples.
This non-zero loss is due to the presence of sinks in our model, which leads to a bias in the assumed outgoing fluxes at the branch points, a phenomenon that should likewise be present in real data.
Thus, unless the dissipation of fluid is taken into account, a direct application of Murray's law cannot be expected to perfectly fit data.

In Fig. \ref{fig:murray}d, where we have non-zero sink fluctuation, the predicted optimal exponents have shifted in the positive direction for all samples.
This shift occurs due to the flattening of the hierarchy of vein widths caused by the fluctuating sinks, and the magnitude of the effect is influenced by both the fluctuating sink amplitude and the fact that the network topologies in leaves from different species are fundamentally different.
The latter results in outgoing veins that are relatively wider than in the case of no sink fluctuation, which in turn implies that the minimum of Eq. \eqref{eq:murray_loss} will occur at a higher value of $\alpha$.
We still observe species-level consistency, as well as greater variability between species than within species, which supports that the predicted exponents are significant.

We now explore the data in more detail by looking at the individual branch nodes. In Fig. \ref{fig:murray}b we plot individual $\super{R_b}{out}$ versus $\super{R_b}{in}$ for all branch nodes in a representative sample of \textit{S. albus},
using the observed optimal value of alpha $\approx 3.03$,
as well as the best-fit value for $\sigma$.
The points very closely follow the line of equality.
This indicates that Murray's law is obeyed over all ranks of veins,
although the inset figure (which shows the same data with logarithmic axes) reveals a larger variation for lower-rank veins, where the effect of sink fluctuation is the highest.
Lastly, Figs. \ref{fig:murray}e-h show residual plots for the same leaf sample and value of $\alpha$.
We use Gaussian weighted means and standard deviations (SD) to illustrate the discrepancies over vein widths.
In Fig. \ref{fig:murray}e, which corresponds to $\sigma = 0$, we see that all residuals are negative, with small variations across vein widths.
The SD is more or less constant across scales, but the residuals also reveal a negative bias.
This suggests that the sink values have to be taken into account in order for Murray's law to be valid for models that employ sinks at the branch nodes.
We investigate this by correcting the residual of each branch node by adding a value proportional to the area of the associated sink.
The resulting residuals can be seen in Fig. \ref{fig:murray}f.
All values have been shifted such that the mean is equal to zero for most branch nodes, while the remaining discrepancies are due to the finite convergence threshold of our simulations.
Figs. \ref{fig:murray}g-h show the corresponding plots for the best-fit value of $\sigma$.
Now, the mean values are positive, and we see a much higher variability, particularly for small-vein branches.
The positive bias occurs due to the aforementioned flattening of the vein width hierarchy.
This means that there are now two effects acting in opposite directions:
The flattening of the vein width hierarchy, and fluid leakage through sinks, which lead to positive and negative bias in the residuals, respectively.
In the same fashion as before, we attempt to correct the residuals by values proportional to the sink areas.
Although the mean values have been shifted towards zero, the effect on the variability is only barely noticeable in this case.

\section{Discussion}
In this paper, we have shown that hydrodynamic models can be run on full-scale leaves, performing analyses on the level of individual network nodes and edges.
This approach stands apart from those of earlier studies and brings with it both advantages and disadvantages.
As an example, looking at the distributions in Figs. \ref{fig:sigma_fit}c-e, it is initially difficult to dismiss the notion that the model faithfully reproduces the image data.
However, in the corresponding plots of individual vein widths in Figs. \ref{fig:sigma_fit}f-h we observe significant deviations from the image data, and even a clear bias in some regimes.
When applied to high-quality data, this aspect of our approach thus provides a more fine-grained method of analysis for leaf venation modeling, in that it exposes the insufficiencies of using mere distributions to compare theoretical venation models to experimental leaf data.

The hydrodynamic model is adapted to the morphology of experimental leaf images by using, e.g., area-weighted, fluctuating sinks.
Including sink fluctuation in the model facilitates the reproduction of loopy structures \cite{katifori2010damage,corson2010fluctuations}, and is in our case controlled by the amplitude parameter $\sigma$.
This parameter was used to perform fits, which yielded species-dependent parameter distributions.
However, the use of a single parameter for this purpose is almost certainly insufficient, as we are fitting tens of thousands of edge widths.
On the other hand, the data is extremely systematic in its variations.
It is therefore pertinent to ask whether the observed discrepancies are caused by either systematic differences or stochastic fluctuations during the venation network formation.
For instance, in our study we set $\gamma = \tfrac{1}{2}$, which defines a volume-based material budget, but this parameter could also turn out to be species-dependent, just as we found for sink fluctuation amplitude.
More generally, being able to compare model and data on an edge-to-edge level allows for an inverse approach to the problem:
instead of merely fitting parameters of predefined models, local feedback models can be parameterized, for example, by neural networks and then optimized.
This would allow data-driven model discovery for leaf venation feedback models.
In particular, we have implemented our approach using \textsc{jax}.
In this study, this has enabled hardware acceleration, but it could further facilitate automatic differentiation in our simulations.
It would therefore be feasible to train such neural networks end-to-end, which is an intriguing avenue for further research.

There is a number of caveats to our approach.
Crucially, we rely on an accurate preprocessing pipeline.
The venation models are compared to the networks extracted from segmented images, thus any bias in this procedure will be reflected in the analyses performed.
Furthermore, we want to demonstrate that both tree-topology networks, as well as fully loopy networks can be reproduced by the model, which implies that output edge widths can take a value of zero.
When mapping the leaf image to the planar graph, we only include the nodes and edges that represent the actual leaf,
so there are, in other words, no edges of zero width in the input data.
Consequently, the model can only be correct if it sets all output edge widths to non-zero values.
According to Figs. \ref{fig:sigma_fit}c-h, however, there is apparently a trade-off:
in order to match the image data, the best-fit model does set a significant number of output edge widths to zero.
This behavior is fundamentally caused by inherent differences between our model and the leaf data.
Conversely, due to the fact that the input network topology is completely determined by the experimental images, we cannot at present observe whether the model would output network edges that do not exist in the input data.
This issue could perhaps be mitigated by adding additional, artificial edges to the leaf input data before running the model, which could further constrain a data-driven model discovery. 

Finally, we showed how our approach enables a novel definition of Murray's law that also applies to loopy networks [Eq. \eqref{eq:murray}].
Thus we circumvent the need to, e.g., transform the graph to an approximate tree-structure \cite{price2013influence}.
We find this definition highly applicable to reticulate networks, although with minor discrepancies, the nature of which we explore further:
When calculating the Murray exponent, we initially assume leak-free flow through the branches,
which contradicts the fact that fluid is uniformly removed from the xylem due to the effects of evaporation and photosynthesis.
Correcting for this, we find that this effect is only important for loopless networks, and the insufficiency of Murray's law on reticulate networks far outweighs the inclusion of fluid dissipation.

We also assume that the reticulate networks are purely optimized for fluid flow, and not for structural support.
This does not happen in plant xylem in general, although the assumption arguably holds for leaves \cite{mcculloh2003water}. 
Additionally, there are indications that the optimization of structural support and fluid flow velocity primarily take place in the larger veins \cite{price2013influence}. 
Since leaves are dominated in number by small veins, structural support can mostly be neglected.

\vspace{1em}
\paragraph{Acknowledgements}
We acknowledge helpful discussions with Louise Isager Ahl. 
This work has received funding from the Novo Nordisk Foundation, Grant Agreement No. NNF20OC0062047.

\bibliographystyle{unsrt}
\bibliography{bibliography}

\end{document}